\documentclass[aps,prl,twocolumn,superscriptaddress,a4paper,
longbibliography, amsmath, amssymb]{revtex4-2}

\usepackage{graphicx}

\begin{document}    

\title{Stellar Temporal Intensity Interferometry\\with Adaptive Optics at the Palomar Observatory}

\author{Darren Ming Zhi Koh}
\affiliation{Centre for Quantum Technologies, 3 Science Drive 2, S117543, Singapore}
\author{Xi Jie Yeo}
\affiliation{Centre for Quantum Technologies, 3 Science Drive 2, S117543, Singapore}
\author{Sam Halverson}
\affiliation{Jet Propulsion Laboratory, California Institute of Technology, 4800 Oak Grove Drive, Pasadena, CA 91109, USA}
\author{Rebecca Oppenheimer}
\affiliation{American Museum of Natural History, Department of Astrophysics, Central Park West at 79th Street, New York, NY 10024, USA}
\author{Gautam Vasisht}
\affiliation{Jet Propulsion Laboratory, California Institute of Technology, 4800 Oak Grove Drive, Pasadena, CA 91109, USA}
\author{Christian Kurtsiefer}
\affiliation{Centre for Quantum Technologies, 3 Science Drive 2, S117543, Singapore}
\affiliation{Department of Physics, National University of Singapore, 2 Science Drive 3, S117542, Singapore}
\author{Peng Kian Tan}
\email{pengkian@gmail.com}
\affiliation{Centre for Quantum Technologies, 3 Science Drive 2, S117543, Singapore}

\date{\today}

\begin{abstract}
Spatial intensity interferometry has seen a major resurgence in astronomy due to its immunity to atmospheric phase turbulence, but its temporal counterpart—a cornerstone of quantum optics—has remained largely unexplored for astronomical sources. 
Here, we demonstrate that high-visibility temporal photon bunching signatures from stellar sources are now accessible by combining ultra-narrowband spectral filtering with high-speed single-photon counting detectors. 
Leveraging the adaptive optics architecture at the Palomar Observatory, we report the first observation of resolved stellar temporal photon bunching, with a peak of $g^{(2)}(\tau=0) = 1.3 \pm 0.1$, from the red supergiant $\alpha$ Her. 
This advance establishes stellar temporal intensity interferometry as a viable observational modality, bridging quantum optics with high-resolution stellar astrophysics.
\end{abstract}

\maketitle

{\em Introduction --}
Spatial intensity interferometry~\cite{hbt:56, hbt:74a} has experienced a renaissance in modern astronomy~\cite{williamson:20, rivet:20, oh:20, pilyavsky:17, pk:16, dravins:15, borra:13, dravins:13, strekalov:13, rou:12, barbieri:09, lebohec:06}, offering phase-independence against atmospheric turbulence and opening pathways toward kilometric-baseline interferometric arrays. 
In contrast, temporal intensity interferometry—a foundational technique in quantum optics~\cite{mandel:95} that directly probes photon coherence in time—has remained virtually unused in optical astronomy for over half a century.

Intensity interferometry relies on measuring higher-order intensity correlations to detect photon bunching.
Thermal light, such as starlight and blackbody radiation, exhibits a characteristic temporal photon bunching behavior~\cite{glauber:63a, glauber:63b, glauber:63c}, known as the Hanbury-Brown--Twiss (HBT) effect~\cite{hbt:56a, hbt:57, hbt:58}. 
This phenomenon is described by the second-order timing correlation function,
\begin{equation}
\label{eqn:g2}
	g^{(2)}(\tau) = 1 + V^{2} \cdot e^{-2|\tau|/\tau_c} \,,
\end{equation} 
which reaches an ideal peak of $g^{(2)}(\tau=0)=2$ for thermal light fields.
The interferometric visibility $V$ scales from 0 to 1 as a measure of experimental imperfections.
Here, $\tau$ represents the timing separation between two photodetection events, and $\tau_c$ parameterizes the coherence timescale, over which thermal photons exhibit a higher probability of arriving in close succession compared to a random Poissonian timing distribution. 
 
The coherence timescale $\tau_{c}$ is inversely proportional to the optical bandwidth, such that $\tau_{c}=1/\Delta f$ for a Gaussian spectrum, where $\Delta f$ is the spectral bandwidth.
Consequently, for starlight with blackbody spectra spanning over hundreds of nanometers, stellar temporal photon bunching exhibits an extremely short coherence timescale $\tau_{c}$ on the order of a few femtoseconds. 

In contrast, the timing jitter $\tau_{j}$ of semiconductor-based single-photon avalanche detectors (SPADs) lies in the nanosecond regime, whereas state-of-the-art superconducting nanowire single-photon detectors (SNSPDs) achieve a few picoseconds. 
Because the thermal coherence time $\tau_{c}$ of starlight is vastly shorter than the detector timing resolution $\tau_{j}$, the photon bunching visibility $V^{2}$ is drastically scaled down by a factor of $\tau_{c}/\tau_{j} \approx 10^{-3}$ in present day literature~\cite{guerin:17}, burying the signal beneath the Poissonian shot-noise background.

In this work, we overcome this fundamental bottleneck to demonstrate that high-visibility temporal photon bunching signatures from starlight are observationally accessible. 
Building upon our earlier demonstration with Sunlight~\cite{pk:14}, we extend this capability to distant stellar targets.
By combining ultra-narrowband spectral filtering with high-speed single-photon detection, alongside the adaptive optics infrastructure at the Palomar Observatory, we detect the first resolved temporal photon bunching signal from an extrasolar star.

{\em Setup --}
\begin{figure}
  \centering\includegraphics[width=\columnwidth]{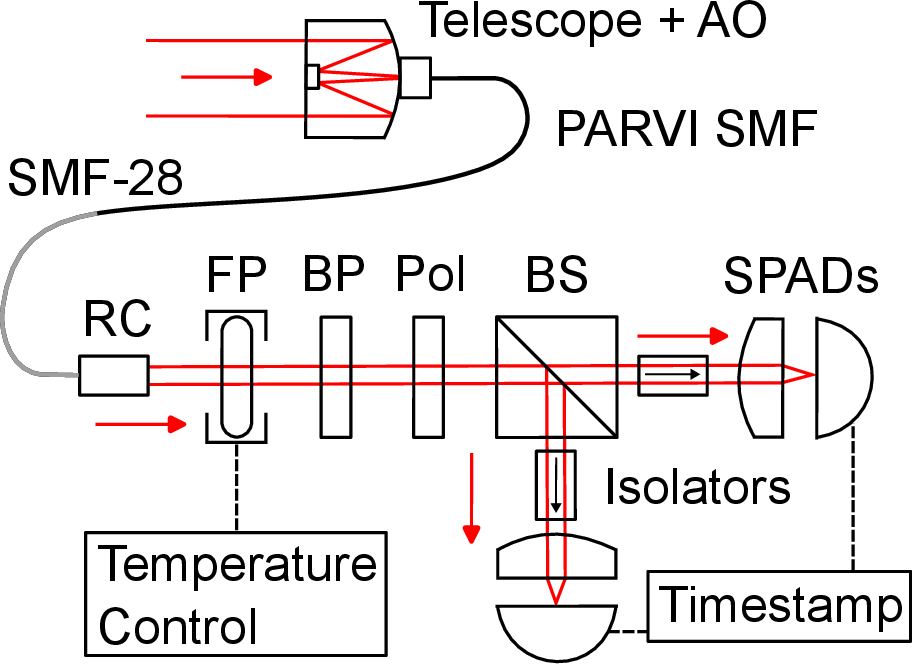}
  \caption{\label{fig:setup} Fiber-coupled starlight is passed through this setup for narrowband spectral filtering. 
The filtered light is then collected by a pair of InGaAs photodetectors and timestamped for timing correlation and data processing. 
AO: adaptive optics, PARVI SMF: PARVI single-mode fiber-feed and transport, SMF28: 1550\,nm single-mode fiber, RC: reflective collimator, FP: Fabry-P\'erot etalon filter, BP: interference bandpass filter, Pol: linear polarization filter, BS: beamsplitter, SPAD: single-photon avalanche detector.}
\end{figure}
Observations were carried out using the 5.1-meter ($f/3.3$) Hale Telescope at the Palomar Observatory. 
To mitigate ground-based seeing limitations, wavefront correction was performed using the PALM-3000 adaptive optics (AO) system~\cite{dekany:13,meeker:20}.
The AO corrected starlight was coupled into the existing 75\,m single-mode fiber-feed and transport of the Palomar Radial Velocity Instrument (PARVI), which interfaces with the PALM-3000 AO~\cite{cale:23}. This infrastructure provides a stable, diffraction-limited optical feed from the Hale Telescope to the downstream instrumentation. AO coupling into single-mode fiber depends on the seeing, which significantly impacted the first two nights of observing. For this present experiment, the fiber output was directed to our dedicated narrowband filtering and photon-correlation apparatus, rather than the PARVI spectrograph, which is fed by a SMF-28 single-mode fiber. The PARVI SMF is fully-compatible with SMF-28 in terms of numerical aperture and core diameter.

The fiber is used to enforce spatial coherence, given that the visibility scales inversely with the number of spatial modes measured. 
Spatial filtering is also essential for optimizing the performance and effective finesse of the downstream spectral filters. 
Optical fibers utilized in this measurement feature angled physical contact (APC) polished tips (at an 8$^{\circ}$ angle) to minimize back-reflections and insertion loss.

The light is collimated to a beam diameter of approximately 2\,mm using an off-axis protected silver parabolic mirror, avoiding chromatic aberration that would otherwise arise from the offset between the measured stellar wavelength (1571.8\,nm) and the alignment laser wavelength (650\,nm) for visual inspection.

To isolate a single temporal mode of thermal light from the star, we employ a combination of a temperature-tuned Fabry-P\'erot (FP) etalon, an interference bandpass (BP) filter, and a linear polarization (Pol) filter (as depicted in Fig.~\ref{fig:setup}). 

\begin{figure}
  \centering\includegraphics[width=\columnwidth]{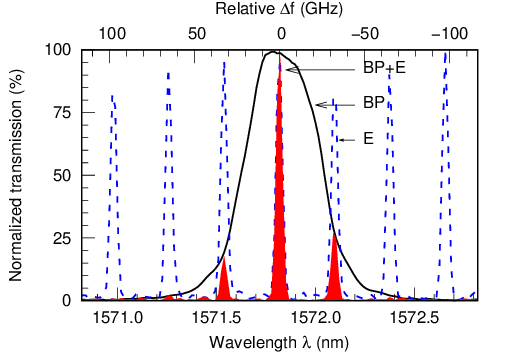}
  \caption{\label{fig:etalon} 
Due to misalignment during transit from Singapore to the observatory, a secondary Fabry-P\'erot etalon (FSR 50\,GHz, passband 4.4\,GHz) was omitted. 
This introduced a spectral mode mismatch between the interference bandpass filter (BP, black solid line), centered at 1571.8\,nm with a 0.45\,nm passband, and the primary Fabry-P\'erot etalon (FP, blue dashed lines) with a finesse of 103, an FSR of about 34.6\,GHz, and a 335\,MHz passband. 
Consequently, secondary side-peaks (red solid peaks) appeared in the transmission spectrum, contributing uncorrelated photoevents and scaling the expected visibility by a factor 0.5, as modeled by the sum of squared probabilities of the intensity ratios of each transmission peak. 
The laboratory thermal light source used a superluminescent diode amplified through an erbium-doped fiber amplifier to provide a continuous broadband spectrum.
Spectrum recorded at 4\,GHz resolution using a laser spectrum analyzer (Bristol 771 series).
}
\end{figure}

The Fabry-P\'erot etalon is built upon a synthetic fused silica (Suprasil 311, Heraeus) substrate. 
Its resonance frequency exhibits a thermal tuning coefficient of about $-1.6$\,GHz/K, governed by the temperature dependence of physical length, and refractive index $n$ described by its Sellmeier coefficients ~\cite{ghosh:97}. 
To model this response, we differentiate the Fabry-Pérot resonance condition for optical frequency $f$ with respect to temperature $T$:
\begin{equation}
\label{eqn:sellmeier}
	\frac{df}{dT} = -f \left( \frac{1}{n}\frac{dn}{dT} + \alpha \right).
\end{equation} 
Using the Sellmeier equations for Suprasil 311, the refractive index at $\lambda = 1571.8\,\text{nm}$ is evaluated as $n = 1.44$. 
Incorporating the material's thermo-optic coefficient ($\frac{dn}{dT} = 1.15 \times 10^{-5}\,\text{K}^{-1}$) and linear thermal expansion coefficient ($\alpha = 5.5 \times 10^{-7}\,\text{K}^{-1}$), we determine an optical frequency shift of approximately $\frac{df}{dT} = -1.62$\,GHz/K (1.3\,pm/K), providing precise thermal tuning control for ultra-narrowband spectral filtering.

\begin{figure}
  \centering\includegraphics[width=\columnwidth]{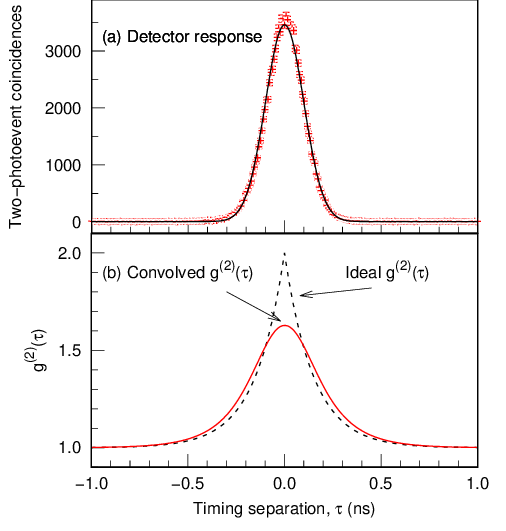}
  \caption{\label{fig:jitter} 
(a) The instrument response function of the detectors is determined by extracting the timing cross-correlation of $\lambda=1550$\,nm photon pairs, which are generated by spontaneous parametric down-conversion (SPDC).
The detectors' combined timing response is well-described by a Gaussian profile with $233.6\pm0.6$\,ps FWHM.
(b) Convolution of the detector timing response with the ideal $g^{(2)}(\tau)$ model (Eqn.~\ref{eqn:g2}) predicts a reduction of visibility $V^{2}$ to 0.6, assuming an ideal thermal state with $V^{2}=1$ and $\tau_{c}=0.3$\,ns.
}
\end{figure} 

Dielectric optical coatings (Laseroptik GmbH) with a reflectivity of $97\,\%$ on both surfaces yield a finesse of 103.
The plano-parallel substrate has a thickness of 3\,mm, corresponding to a free spectral range (FSR) of approximately 34.6\,GHz (0.285\,nm), with a periodic transmission window of 335\,MHz (2.77\,pm) full width at half maximum (FWHM). 
The interference bandpass (BP) filter provides a 0.45\,nm (55\,GHz) wide passband centered at $\lambda = 1571.8$\,nm to suppress background stellar light outside the etalon's free spectral range.

The spectrally filtered starlight passes through a linear polarization filter to maximize bunching contrast; orthogonal polarization states are mutually incoherent and thus contribute exclusively to the uncorrelated Poissonian baseline of $g^{(2)}(\tau) = 1$. 
The beam is then split via a 50:50 non-polarizing beamsplitter, with each output routed through an optical isolator to suppress detector-induced breakdown flashes~\cite{kurtsiefer:01a, yicheng:17} that can generate spurious correlated photoevents.

Photo-detection is performed by a pair of actively quenched Indium Gallium Arsenide (InGaAs) single-photon avalanche detectors (SPADs) featuring a quantum efficiency of approximately $30\,\%$ at 1550\,nm and a combined timing jitter FWHM of roughly 230\,ps (Fig.~\ref{fig:jitter}). 
The instrument response function was characterised by determining the timing cross-correlation between the two output beams generated by spontaneous parametric down conversion (SPDC)~\cite{lijiong:21, lijiong:22}. 
The detectors were configured with a 10\,$\mu$s dead-time to mitigate afterpulsing effects that could otherwise bias timing correlations.
Detection timestamps are recorded by a time-to-digital converter with a 3.6\,ps resolution and subsequently binned into 128\,ps intervals to compute the second-order temporal correlation function $g^{(2)}(\tau)$.

\begin{figure}
  \centering\includegraphics[width=\columnwidth]{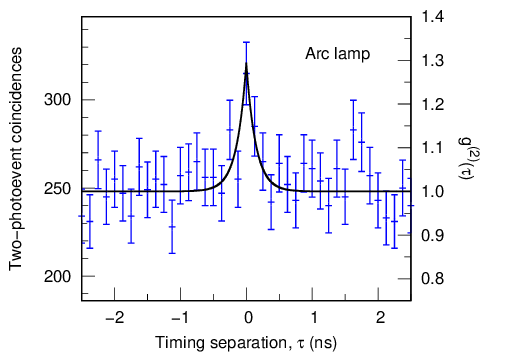}
  \caption{\label{fig:arclamp}
Baseline $g^{(2)}(\tau)$ measurements acquired using an Argon arc lamp (color temperature $T = 6000\,\text{K}$) as a reference laboratory thermal source.
The measured photon bunching is $g^{(2)}(0) = 1.3 \pm 0.1$ with a coherence timescale of $\tau_{c} = 0.3 \pm 0.1\,\text{ns}$, fitted to a reduced $\chi^{2}/$doF$=0.998$.
This reduction in the visibility $V^{2}$ from theory (Eqn.~\ref{eqn:g2}) is consistent with a factor of 0.6 arising from the detector timing response (Fig.~\ref{fig:jitter}), combined with a factor of 0.5 due to the etalon optical response (Fig.~\ref{fig:etalon}).
}
\end{figure}

\begin{figure}
  \centering\includegraphics[width=\columnwidth]{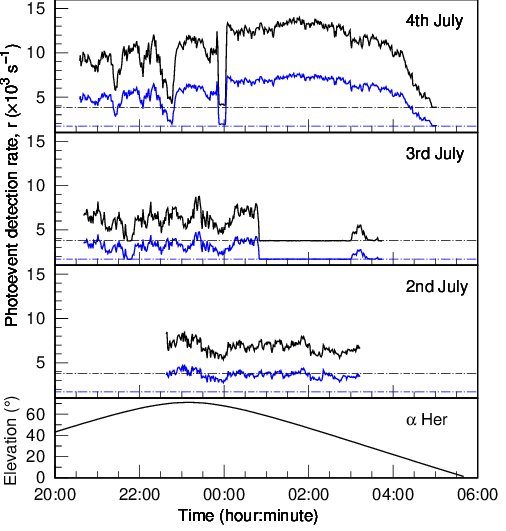}
  \caption{\label{fig:countrate} Photoevent detection rates, $r$, recorded over 3 nights of observation from the 2nd to 5th of July 2026. 
Observation windows spanned 8:30\,pm to 5:30\,am local time, with peak performance achieved by the adaptive optics system when the target star, $\alpha$ Her, approached transit or maximal elevation angle near midnight.
The adaptive optics system underwent a prolonged reboot cycle during the second night, which led to the detection rates dropping to detector dark count levels (dashed horizontal lines) from 1:00\,am to 3:00\,am.
The InGaAs SPAD channels exhibited slight differences in quantum efficiency and dark count rates, resulting in  an offset between detection rates as indicated by the black and blue traces.
}
\end{figure}

{\em Results --}
Observations were conducted over three nights at the 5.1-meter Hale Telescope at the Palomar Observatory from the 2nd to 4th July, 2026 (latitude: 33$^{\circ}$\,21$^{\prime}$\,23$^{\prime\prime}$\,N, longitude: 116$^{\circ}$\,51$^{\prime}$\,54$^{\prime\prime}$\,W, elevation: 1712\,m). 
The photoevent detection rates $r$ are shown in Fig.~\ref{fig:countrate}, with individual detector dark count rates of approximately 3800\,photoevents s$^{-1}$ and 1700\,photoevents s$^{-1}$ respectively.

\begin{figure}
  \centering\includegraphics[width=\columnwidth]{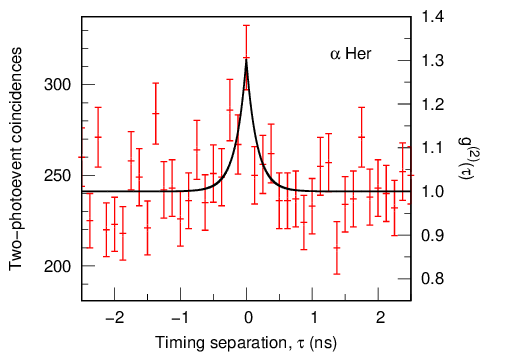}
  \caption{\label{fig:g2} Stellar temporal intensity interferometric measurement $g^{(2)}(\tau)$ of $\alpha$ Her, a red supergiant located over 360\,light-years away, with a peak photon bunching of $g^{(2)}(\tau=0) = 1.3\pm0.1$ and a coherence timescale of $\tau_{c}=0.3\pm0.1$\,ns, fitted at $\chi^{2}/\textrm{doF}=0.998$.
The observation was integrated over 7 hours during the night of 4th/5th July 2026, with a signal detection rate of around 7000 photoevents s$^{-1}$ per detector.
}
\end{figure}

Our stellar target was $\alpha$ Her (Rasalgethi), a hierarchical triple star system featuring a red supergiant primary, located at a distance of roughly 360\,light-years. 
Its apparent magnitude in the H-band (1650\,nm) is $-3.22$ (SIMBAD). 
Detections during the first two nights were relatively noisy, with the stellar signal rates close to the respective detector dark count rates.
Thus only the data collected for the third night (4th-to-5th July) were processed, rejecting the timebins with detection rates below the dark count threshold, and computed into the second-order timing correlation function $g^{(2)}(\tau)$ as shown in Fig.~\ref{fig:g2}.

With this setup, we observed photon bunching in light from $\alpha$ Her, with a peak of $g^{(2)}(\tau=0) = 1.3\pm0.1$ and a coherence timescale of $\tau_{c}=0.3\pm0.1$\,ns.
This is in good agreement with theory (Eqn.~\ref{eqn:g2}), after accounting for both the optical response of the etalon spectral mode mismatch (Fig.~\ref{fig:etalon}), with the electronic response of the detectors' timing jitter (Fig.~\ref{fig:jitter}), and referenced against the laboratory thermal light source (Fig.~\ref{fig:arclamp}).

{\em Signal-to-noise consideration --}
When photodetectors are fast enough to resolve the temporal coherence $\tau_{c}$ of the photon bunching signature, the signal-to-noise ratio (SNR) of the second-order correlation function $g^{(2)}(\tau)$ will be dominated by shot noise of the photodetection events, and can be described by \cite{hbt:74, foellmi:09}
\begin{equation}
\label{eqn:snr}
	\mathrm{SNR} = r \cdot V^{2} \sqrt{\tau_{c} \cdot \Delta T} \,,
\end{equation} 
with the photoevent rate $r$, the interferometric visibility $V=\sqrt{g^{(2)}(0)-1}$, the coherence time $\tau_{c}$, and the integration time $\Delta T$.
Using the measured photon bunching parameters ($V^2 = 0.3$ and $\tau_c = 0.3$\,ns from Fig.~\ref{fig:g2}), an SNR of approximately 3 is reached after an integration time of $\Delta T = 7$\,hours at a detection rate of $r = 7000$\,s$^{-1}$.

A natural path forward for scaling this technique involves upgrading from InGaAs single-photon avalanche detectors (SPADs) to superconducting nanowire single-photon detectors (SNSPDs). 
Implementing SNSPDs would increase quantum efficiency (QE) from $\sim 30\,\%$ to over $90\,\%$, while reducing the dark count rate from several 1000\,s$^{-1}$ down to sub-$100$\,s$^{-1}$. 
Furthermore, the reduced timing jitter (from $\sim 300$\,ps down to a few tens of picoseconds) would enable the use of a wider spectral passband ($\sim 1$\,GHz), substantially increasing the signal detection rate. 

\begin{figure}
  \centering\includegraphics[width=\columnwidth]{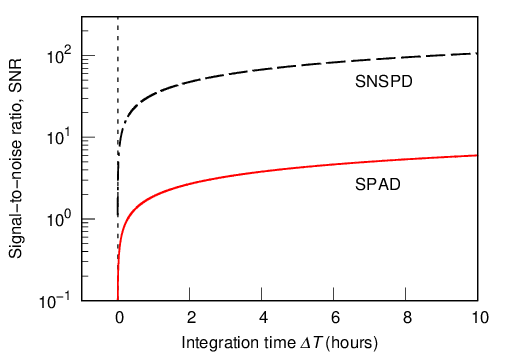}
\caption{\label{fig:snr} Projected signal-to-noise ratio comparing future implementation using superconducting nanowire single-photon detectors (SNSPDs, black dashed line) against the current InGaAs SPAD configuration (red solid line), under identical observational conditions.
}
\end{figure}

Combined with corrections to the etalon mode mismatch by employing a thinner substrate (1\,mm) to yield a free spectral range wider than the 0.45\,nm filter passband, these upgrades will enhance the resolved visibility from $V^{2}=0.3$ toward ideal limits and improve overall signal acquisition rates by an order of magnitude, as illustrated in Fig.~\ref{fig:snr}.

{\em Summary --}
We have demonstrated the first high-visibility observation of temporal photon bunching from an extrasolar celestial source, measuring $g^{(2)}(\tau=0) = 1.3 \pm 0.1$ from the red supergiant $\alpha$ Her. 
By combining high-speed adaptive optics at the Palomar Observatory with ultra-narrowband spectral filtering and single-photon counting detectors, we overcome the severe temporal resolution bottlenecks that have long restricted intensity interferometry to the spatial domain for 70\,years.

This milestone opens an exciting observational regime for quantum astronomy, bridging high-resolution astronomical imaging with quantum optics, and establishing temporal intensity interferometry as a robust tool for astrophysical research. 
Beyond fundamental stellar characterization, this technique enables the investigation of non-thermal emission~\cite{xijie:24, xijie:26} from natural astrophysical laser~\cite{dravins:08} candidates—such as Eta Carinae~\cite{johansson:05} and Wolf-Rayet stars like Gamma Velorum~\cite{hbt:70}—as well as the search for optical technosignatures~\cite{siemion:23}. 
Furthermore, high-precision temporal correlation spectroscopy~\cite{pk:17} grants access to ultra-narrow spectral features inaccessible to conventional dispersive optics, while offering unique avenues to constrain cosmic-scale timing dispersion~\cite{milburn:91}, and test models of Lorentz invariance violation~\cite{milburn:06}.

\begin{acknowledgments}
This research is supported by the Quantum Engineering Programme through
grants QEP-P1 and NRF2021-QEP2-03-P02, the National Research Foundation, Prime
Minister's Office, Singapore.

This work was based on observations obtained at the Hale Telescope, Palomar Observatory, as part of a collaborative agreement between the Caltech Optical Observatories and the Jet Propulsion Laboratory (operated by Caltech for NASA).
We thank the Palomar Observatory telescope operators, support astronomers, hospitality, and administrative staff, without whom this research would not have
been possible.
\end{acknowledgments}

%

\end{document}